\documentclass[aps,pra,superscriptaddress,onecolumn,nofootinbib,a4paper]{revtex4}

\usepackage{graphicx,graphics,epsfig}   % Include figure files
\usepackage{dcolumn}    % Align table columns on decimal point
\usepackage{bm}         % bold math
\usepackage{verbatim}   % useful for program listings
\usepackage{color}      % use if color is used in text
\usepackage{subfigure}  % use for side-by-side figures
\usepackage{times}
\usepackage[normalem]{ulem}

\usepackage{amsmath}    % need for subequations
\usepackage{amssymb}

\usepackage{footmisc}

\newcommand{\be}{\begin{equation}}
\newcommand{\ee}{\end{equation}}
\newcommand{\ba}{\begin{eqnarray}}
\newcommand{\ea}{\end{eqnarray}}
\newcommand{\ban}{\begin{eqnarray*}}
\newcommand{\ean}{\end{eqnarray*}}

\newcommand{\ket}[1]{\mbox{$ | #1 \rangle $}}
\newcommand{\bra}[1]{\mbox{$ \langle #1 | $}}

\newcommand{\one}{\leavevmode\hbox{\small1\normalsize\kern-.33em1}}

\definecolor{nblue}{rgb}{0.2,0.2,0.7}
\definecolor{ngreen}{rgb}{0.2,0.6,0.2}
\definecolor{nred}{rgb}{0.8,0.2,0.2}
\definecolor{nblack}{rgb}{0,0,0}

\begin{document}

\title[Device-dependent and device-independent QKD without a shared reference frame]{Device-dependent and device-independent quantum key distribution without a shared reference frame}

\author{Joshua A. Slater}
\affiliation{Institute for Quantum Science and Technology, University of Calgary, Canada}
\author{Cyril Branciard}
\affiliation{School of Mathematics and Physics, The University of Queensland, St Lucia, QLD 4072, Australia}
\author{Nicolas Brunner}
\affiliation{D\'epartement de Physique Th\'eorique, Universit\'e de Gen\`eve, 1211 Gen\`eve, Switzerland}
\affiliation{H.H. Wills Physics Laboratory, University of Bristol, Tyndall Avenue, Bristol, BS8 1TL, United Kingdom}
\author{Wolfgang Tittel}
\affiliation{Institute for Quantum Science and Technology, University of Calgary, Canada}

\date{\today}

\begin{abstract}
Standard quantum key distribution (QKD) protocols typically assume that the distant parties share a common reference frame. In practice, however, establishing and maintaining a good alignment between distant observers is rarely a trivial issue, which may significantly restrain the implementation of long-distance quantum communication protocols. Here we propose simple QKD protocols that do not require the parties to share any reference frame, and study their security and feasibility in both the usual device-dependent case---in which the two parties use well characterized measurement devices---as well as in the device-independent case---in which the measurement devices can be untrusted, and the security relies on the violation of a Bell inequality. To illustrate the practical relevance of these ideas, we present a proof-of-principle demonstration of our protocols using polarization entangled photons distributed over a coiled 10-km-long optical fiber. We consider two situations, in which either the fiber spool freely drifts, or randomly chosen polarization transformations are applied. The correlations obtained from measurements allow, with high probability, to generate positive asymptotic secret key rates in both the device-dependent and device-independent scenarios (under the fair-sampling assumption for the latter case).
\end{abstract}

\maketitle

\section{Introduction}

Quantum Key Distribution \cite{QKDreview} is arguably the most developed area in quantum information processing, and has recently reached the commercial level. Currently the main limitation of QKD is the distance between the parties. State of the art experiments have reported key exchanges up to distances of $\sim 250$ km~\cite{Stucki2009}. It is a great challenge in this area to reach much longer distances, such as intercontinental distances, and tremendous effort is made in this direction. Significant progress has been recently reported, with promising developments in quantum repeaters \cite{Qrep}, as well as in satellite-based quantum communications~\cite{satellite,Bourgoin13}. 

The main reason for this challenge to practical implementations of QKD protocols, and more generally all long-distance quantum communication tasks, is the effect of noise and loss. Many studies have been devoted to these problems. There is however another key issue, often overseen, which is the alignment of a common reference frame between the parties. While usually being assumed {\it a priori} (hence not discussed) in theoretical works, the alignment of a common reference frame is rarely a trivial task in practice. Furthermore, when performing experiments outside of the laboratory, this issue can become highly cumbersome, and may significantly restrain---and even hinder---the implementation of certain quantum protocols. For instance, in fiber-based quantum communications, polarization rotations are induced by unavoidable temperature changes, which makes it challenging to maintain a good alignment. Also, in satellite-based quantum communications, establishing and maintaining a good alignment between the satellite and the ground station is a challenge~\cite{Bourgoin13}, given the fast movement of the satellite and the limited amount of time for completing the protocol. 

It is therefore relevant to consider quantum communication protocols in which the requirement of a common reference frame can be dispensed with. An elegant solution to this problem is to use decoherence-free subspaces \cite{cabello,spekkens}. However, this generally amounts to using high-dimensional quantum systems, the practical implementation of which is challenging---although progress has been achieved recently \cite{ambrosio}.
It turns out however that one can in fact relax the shared reference frame assumption in certain simple quantum communication protocols that only involve qubits. This approach has received some attention in the context of tests of quantum nonlocality \cite{liang}: in particular it was recently shown~\cite{Shadbolt,wallman}, and experimentally illustrated~\cite{Shadbolt,palsson}, that Bell inequality violations can be guaranteed even if the parties share no common measurement basis. In the context of QDK, Laing and colleagues \cite{laing} presented a protocol---dubbed ``Reference Frame Independent'', and recently implemented in Ref. \cite{wabnig}---which requires the parties to only have one common measurement basis. 
While the latter approach is well suited and proposes an interesting solution for certain QKD implementations, it is however not adapted to all systems.	

Here we propose QKD protocols that do not assume the existence of any shared reference frame. We analyse their security and feasibility in two scenarios. In the first, ``device-independent'' (DI) case~\cite{MayersYao,DI_PRL}, the two communicating parties Alice and Bob use untrusted measurement devices and do not make any assumption on their functioning; the security of the protocol is ensured by the violation of a Bell inequality (for a recent review, see~\cite{Brunner13}). In the second, standard ``device-dependent'' (DD) case, Alice and Bob trust that their devices faithfully implement the prescribed measurements---which further constrains the possible attacks by an eavesdropper, Eve, detectable by Alice and Bob.
We show in both cases that if Alice and Bob do not share a common reference frame but measure entangled pairs of quantum systems along randomly orientated measurement bases, they can still expect to generate, with reasonably large probability (which depends on the assumptions for the security analysis), secret keys with positive key rates.
We then demonstrate the experimental relevance of these ideas by presenting a proof-of-principle implementation of our protocols using a photonic QKD setup with polarization entangled photons. For all cases under consideration, we could calculate, with non-zero probability, positive (asymptotic) secret key rates as obtained from the preceding security analysis (assuming the fair sampling assumption in the DI case to calculate the violation of a Bell inequality). This suggests that the requirement of a common reference frame can indeed---if need be---be completely dispensed with in experimental QKD, thus opening promising perspectives for long-distance and satellite-based QKD.

\section{Device-independent protocol}\label{sec:DI}

In our first protocol, Alice and Bob can each perform one out of 3 possible local measurements, labeled by $x = 1,2,3$ for Alice and $y = 1,2,3$ for Bob, on a shared entangled quantum state $\rho_{AB}$. All measurements are dichotomic, giving a binary outcome $a$ for Alice and $b$ for Bob. The protocol is device-independent in the sense that we shall not make any assumption on which measurements are physically implemented by Alice and Bob's measuring apparatuses, nor of the dimension of the state $\rho_{AB}$.

After repeating the above operations sufficiently many times,  Alice and Bob can, by communicating a random subset of their measurement choices and results, estimate the correlations they share, i.e. the probability distribution $P(a,b|x,y)$.
For now we will focus on the {\it correlators} $E_{xy} = P(a{=}b|x,y) - P(a{\neq}b|x,y)$. 
From these, Alice and Bob can in particular calculate the 36 values (for all $x, x', y$ and $y'$) of the Clauser-Horne-Shimony-Holt (CHSH)~\cite{chsh} parameters
\ba S_{xx'yy'} = |E_{xy} + E_{xy'} + E_{x'y} - E_{x'y'}|. \label{eq:def_chsh} \ea
If any of these CHSH values is greater than 2, Alice and Bob can certify that the observed correlations are ``nonlocal'', in the sense that they violate Bell's local causality assumption~\cite{bell_book}.
Observing quantum nonlocality is not only interesting for testing the foundations of quantum theory, it can also have practical applications---in our case of interest it can indeed allow one, for a large enough value of a CHSH parameter together with a large enough value for at least one correlator, to prove the security of QKD protocols in a device-independent way~\cite{MayersYao,DI_PRL,DI_NJP,DI_NatComm,hanggi_renner}.

Interestingly, it was shown in Refs.~\cite{Shadbolt,wallman} that if Alice and Bob share a maximally entangled 2-qubit state, say the singlet state $\ket{\Psi^-} = \frac{1}{\sqrt{2}} (\ket{01}-\ket{10})$,  and can each choose among 3 orthogonal measurements, represented for Alice (Bob) by 3 orthogonal vectors $\vec a_x$ ($\vec b_y$) on the Bloch sphere, there is always at least one of the 36 CHSH values $S_{xx'yy'}$ that is above the local bound of 2---unless Alice and Bob's orthogonal measurement triads are perfectly aligned. Moreover, if Alice and Bob do not share any common reference frame and the relative orientation of their measurement triads is random, the largest CHSH value they observe is typically quite large: its average value was empirically found to be $\sim 2.6$ for random relative orientations drawn from a uniform distribution on the Bloch sphere~\cite{Shadbolt}.

This suggests that it should be possible to extract reasonably large secret key rates from the correlations obtained by Alice and Bob, without the requirement that they share a common reference frame (i.e. their orthogonal measurement triads are not pre-aligned). We study this idea below, following the security analyses of both Pironio \emph{et al.}~\cite{DI_NJP}---which proves the security against \emph{collective attacks}---and of Masanes \emph{et al.}~\cite{DI_NatComm}---which considers the security against \emph{general (coherent) attacks}, only assuming \emph{memoryless devices}. Note that full security proofs of DI QKD, considering the most general attacks, were recently reported~\cite{Reichardt13,Vazirani12}. However, these proofs are not robust to noise, hence of limited practical interest, and we do not consider these in this work.

\subsection{Device-independent security analysis along the lines of Pironio \emph{et al.}~\cite{DI_NJP}}

Ref.~\cite{DI_NJP} considered a DI-QKD protocol with 3 inputs for Alice (in our notations, $x = 1,2,3$) and 2 inputs for Bob ($y = 1,2$). Considering the CHSH parameter $S_{1212}$ and the correlator $E_{31}$ (see Eq.~\ref{eq:def_chsh} above), it was shown that (if $S_{1212} > 2$) a secret key can be extracted through 1-way classical post-processing (from Bob to Alice) from the data obtained when using the settings $x = 3$ and $y = 1$---the ``raw key''---at an asymptotic rate (see details in~\cite{DI_NJP})
\ba
R \ \geq \ 1 - h \Big[ \frac{1-E_{31}}{2} \Big] - h \Big[ \frac{1 + \sqrt{(S_{1212}/2)^2-1}}{2} \Big], \label{eq:R_DI_NJP}
\ea
where $h(p) = -p \log_2 p - (1{-}p) \log_2 (1{-}p)$ is the binary entropy function.
This bound on the secret key-rate ensures the security of the QKD protocol in the DI scenario against \emph{collective attacks}~\cite{QKDreview}, in the limit of infinite key lengths.
The term $h \big[ \frac{1-E_{31}}{2} \big]$ represents the (minimum) amount of information that Alice and Bob need to classically exchange in order to correct the errors in their raw keys, while the term $h \big[ \frac{1 + \sqrt{(S_{1212}/2)^2-1}}{2} \big]$ is a bound on Eve's Holevo information conditioned on Bob's measurement result. Both need to be reduced through privacy amplification.

The same protocol as in~\cite{DI_NJP} can be run by following the protocol detailed above, in which Alice and Bob do not share a common reference frame (see start of Sec.~\ref{sec:DI}), with 3 settings for both Alice and Bob, by using any four correlators $E_{x^{(\prime)}y^{(\prime)}}$ to estimate a CHSH parameter $S_{xx'yy'}$, and any pair of settings $(x_{\text{raw}}, y_{\text{raw}})$ to define the raw key---with the important condition that either $x_{\text{raw}} \in \{x,x'\}$ or $y_{\text{raw}} \in \{y,y'\}$.
Let us then define
\ba
\hspace{-20mm} r_{DI_1} = \max_{\stackrel{x,x',y,y',x_{\text{raw}},y_{\text{raw}},}{\stackrel{\text{s.t. } \ S_{xx'yy'} > 2,}{x_{\text{raw}} \in \{x,x'\} \ {\mathrm{or}} \ y_{\text{raw}} \in \{y,y'\}}}} \Bigg[ 1 - h \Big[ \frac{1{-}E_{x_{\text{raw}} y_{\text{raw}}}}{2} \Big] - h \Big[ \frac{1{+}\sqrt{(S_{xx'yy'}/2)^2{-}1}}{2} \Big] \Bigg] , \quad \label{eq_rDI_1}
\ea
corresponding to the rate~\eqref{eq:R_DI_NJP} for the optimal choice of settings used to define the CHSH parameter and the raw key (by convention, if no violation $S_{xx'yy'} > 2$ is found we define $r_{DI_1} = -1$. 
From the analysis of~\cite{DI_NJP}, if $r_{DI_1}$ is found to be non-negative, then Alice and Bob will indeed be able to extract a secret key with an asymptotic rate of (at least) $r_{DI_1}$, in the DI scenario, secure against collective attacks\footnote{Note that instead of throwing some raw data away, Alice and Bob could additionally try to extract some secret key from their measurement results obtained by using other settings than the optimal $(x_{\text{raw}}, y_{\text{raw}})$, if any other choice also leads to a positive bound on the key rate through~\eqref{eq:R_DI_NJP}. For simplicity we do not consider this possibility in this paper, and only focus on $r_{DI_1}$ as defined in~\eqref{eq_rDI_1}.}.

\medskip

In order to study the experimental feasibility of such a protocol the questions we need to address are the following: How likely is $r_{DI_1}$ to be positive? What are its typical values, and how are they distributed if Alice and Bob's orthogonal measurement triads are randomly chosen?

\begin{figure}%[h!]
\centering
    \includegraphics[width=8cm]{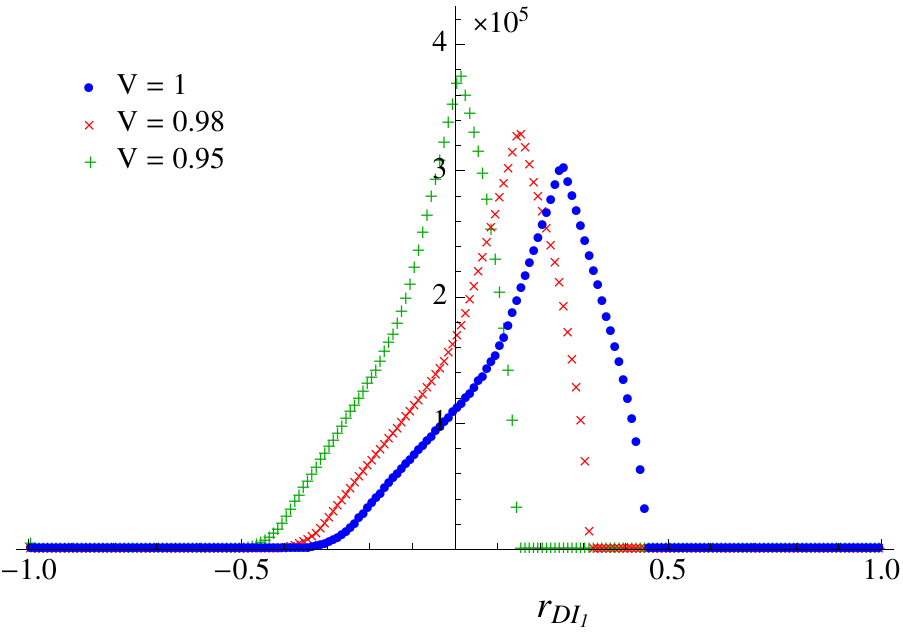}
    \caption{Estimated distribution of the bound $r_{DI_1}$~(\ref{eq_rDI_1}) on the secret key rate (in the DI scenario, following the analysis of~\cite{DI_NJP}---i.e. ensuring security against collective attacks), obtained by generating $10^7$ random pairs of orthogonal measurement triads uniformly distributed on the Bloch sphere, to be measured on Werner states of visibilities $V = 1, 0.98$ and 0.95. For each value of $V$, each data point corresponds to the number of samples (out of $10^{7}$) giving a value $r_{DI_1}$ within an interval of sive $\delta r = 0.01$.}
    \label{fig_distribution_rDI_1}
\end{figure}

To answer these questions, we estimated the distribution of $r_{DI_1}$ by generating $10^7$ pairs of random orthogonal measurement triads $\{\vec a_x\}$ and $\{\vec b_y\}$, independently drawn from a uniform distribution on the Bloch sphere. For each pair of triads, we computed the 9 correlators $E_{xy}$ assuming that Alice and Bob receives pure (noise-free) maximally entangled states $\rho_{AB} = \ket{\Psi^-}\!\bra{\Psi^-}$ (hence, $E_{xy} = -\vec a_x \cdot \vec b_y$), and calculated the bound $r_{DI_1}$~(\ref{eq_rDI_1}) on the secret key rate that Alice and Bob can extract. The results of our simulation are plotted on Figure~\ref{fig_distribution_rDI_1}. We found that $\sim 83.9\%$ of our samples of $r_{DI_1}$ were positive (i.e., we estimate the probability for Alice and Bob to obtain $r_{DI_1}>0$ to be $\sim 83.9\%$), with a maximum value for the distribution of $r_{DI_1}$ observed around $r_{DI_1} \sim 0.25$. The average value of $r_{DI_1}$ was found to be $\sim 0.173$; if we post-select only the cases in which $r_{DI_1}>0$, the average value becomes $\sim 0.226$.
The maximum value for $r_{DI_1}$ is found to be $\sim 0.450$, obtained if two of Alice and Bob's measurement settings coincide (say, $\vec a_{x''}=\vec b_{y''}$), while the other two pairs of settings, used to define $S_{xx'yy'}$, are coplanar (with an angle $\sim 0.642$ rad from one pair to the other).

It is also important to study the effect of noise on the secret key rates $r_{DI_1}$. For that, we similarly estimated the distribution of $r_{DI_1}$ if the measurements are now performed on noisy singlet states (Werner states~\cite{werner89}) $\rho_{AB}^V = V \ket{\Psi^-}\!\bra{\Psi^-} + (1{-}V)\one/4$ (which gives $E_{xy} = - V \vec a_x \cdot \vec b_y$), for $V = 0.98$ and 0.95; see Figure~\ref{fig_distribution_rDI_1}. As expected, the secret key rates are reduced as $V$ decreases. For $V = 0.98$ and $V = 0.95$, the probabilities that $r_{DI_1}>0$ are, however, still $\sim 72.1 \%$ and $\sim 38.0 \%$, respectively. Note in this respect that the violations of a CHSH inequality were found in Ref.~\cite{Shadbolt} to be quite robust to noise; for instance, the probability that at least one value of $S_{xx'yy'}$ is greater than 2 is still above $99.9 \%$ for $V = 0.95$.

\subsection{Device-independent security analysis along the lines of Masanes \emph{et al.}~\cite{DI_NatComm}}

Ref.~\cite{DI_NatComm} provides a different approach to prove the security of a DI-QKD scheme. For the same protocol as in Ref.~\cite{DI_NJP}, Masanes \emph{et al.} proved that (for $S_{1212} > 0$) a secret key---now secure against {\it coherent attacks, but under the assumption that their measurement devices are causally independent} (or \emph{memoryless})---can be extracted at an asymptotic rate~\cite{DI_NatComm}
\ba
R \ \geq \ - \ h \Big[ \frac{1-E_{31}}{2} \Big] - \log_2 \Big[ \frac{1 + \sqrt{2-(S_{1212}/2)^2}}{2} \Big].
\ea
Again, the term $h \big[ \frac{1-E_{31}}{2} \big]$ is due to the necessary error correction, while the term $\log_2 \big[ \frac{1 + \sqrt{2-(S_{1212}/2)^2}}{2} \big]$ is now a bound on the min-entropy of Alice's raw key conditioned on Eve's information. The information of both must be removed through privacy amplification to extract a secret key.

Let us then now define, for our experimental procedure with 3 settings for Alice and Bob,
\ba
\hspace{-22mm} r_{DI_2} = \! \max_{\stackrel{x,x',y,y',x_{\text{raw}},y_{\text{raw}},}{\stackrel{\text{s.t. } \ S_{xx'yy'} > 2,}{x_{\text{raw}} \in \{x,x'\} \ {\mathrm{or}} \ y_{\text{raw}} \in \{y,y'\}}}} \! \Bigg[ - h \Big[ \frac{1{-}E_{x_{\text{raw}} y_{\text{raw}}}}{2} \Big] - \log_2 \Big[ \frac{1 + \sqrt{2{-}(S_{xx'yy'}/2)^2}}{2} \Big] \Bigg] . \label{eq_rDI_2}
\ea
As before, if $r_{DI_2}$ is found to be non-negative, then Alice and Bob will indeed be able to extract a secret key with a rate (at least) $r_{DI_2}$.

\medskip

\begin{figure}%[h!]
\centering
    \includegraphics[width=8cm]{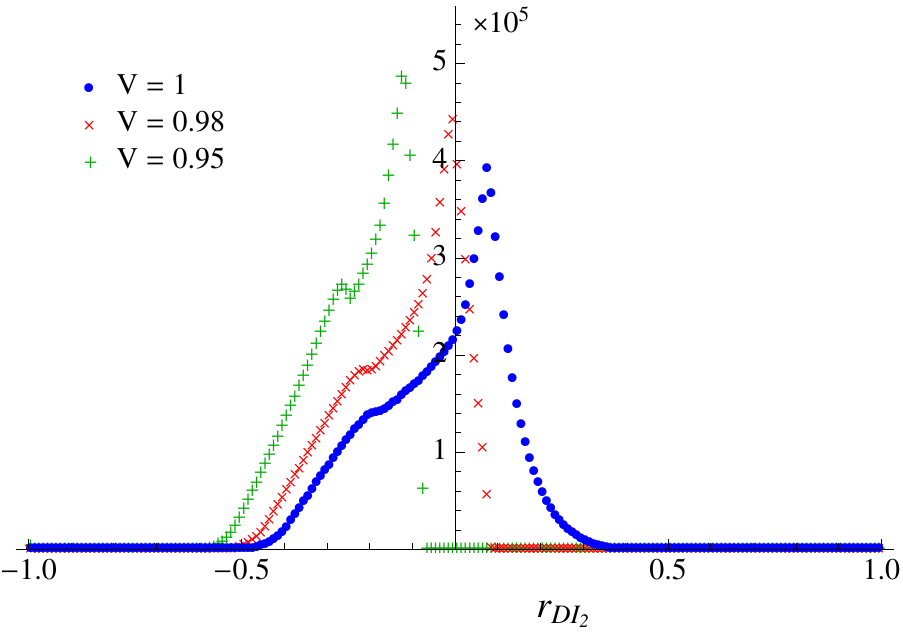}
    \caption{Estimated distribution of the bound $r_{DI_2}$~(\ref{eq_rDI_2}) on the secret key rate for Werner states of visibilities $V = 1, 0.98$ and 0.95 (in the DI scenario, following the analysis of~\cite{DI_NatComm}---i.e. showing security against coherent attacks, for memoryless devices), obtained as in Figure~\ref{fig_distribution_rDI_1}.}
    \label{fig_distribution_rDI_2}
\end{figure}

Again, we wish to determine how likely it is that $r_{DI_2}$ is positive, and how its typical values and distribution look like when Alice and Bob's orthogonal measurement triads are randomly chosen from a uniform distribution on the Bloch sphere.
For that, we estimated the distribution of $r_{DI_2}$ in a similar way as for $r_{DI_1}$. The results of our simulation are plotted in Figure~\ref{fig_distribution_rDI_2}. We found that $\sim 49.0\%$ of our samples of $r_{DI_2}$ were positive, and observed a peak in the distribution of $r_{DI_2}$ for values around $\sim 0.08$. The average value of $r_{DI_2}$ was found to be $\sim -0.034$; if we post-select only the cases in which $r_{DI_2}>0$, the average value becomes $\sim 0.093$.
The maximum value for $r_{DI_2}$ is obtained if two of Alice and Bob's measurement settings coincide, while the other two pairs of settings, used to define $S_{xx'yy'}$, are coplanar, at $45^\circ$ from one another---i.e. they correspond to the optimal choice of settings for testing the CHSH inequality (which was not the case for the optimal settings for $r_{DI_1}$). In that case, one gets $r_{DI_2} = 1 - h(\frac{1-1/\sqrt{2}}{2}) \simeq 0.399$.

We also, as before, considered the effect of depolarizing noise on the secret key rates $r_{DI_2}$ (see Figure~\ref{fig_distribution_rDI_2}). For $V = 0.98$, we found the probability that $r_{DI_2}>0$ to be $\sim 18.0 \%$; for $V = 0.95$, however, no positive secret key rate $r_{DI_2}$ is obtained any more.

Note that $r_{DI_2}$ is always smaller than $r_{DI_1}$. This comes from the different techniques used in the proofs: Ref.~\cite{DI_NatComm} is based on the calculation of min-entropies to estimate Eve's information, while Ref.~\cite{DI_NJP} is based on the calculation of Eve's Holevo information (which involves von Neumann entropies). The security analysis of Ref.~\cite{DI_NatComm} is more stringent in that it considers more general attacks.
It is an open question whether any of the two analyses can be improved to account for more general attacks or to lead to higher bounds on the secret key rates (e.g. whether the higher bound of Ref.~\cite{DI_NJP} also holds for the same class of attacks as considered in Ref.~\cite{DI_NatComm}).

\section{Device-dependent protocol}\label{sec:DD}

We now turn to the more standard device-dependent (DD) scenario, in which Alice and Bob trust their measurement apparatuses. We assume that the apparatuses implement dichotomic \emph{qubit} measurements, that $\rho_{AB}$ is a 2-qubit state, and that the 3 measurement settings they can each choose from, as before, trustfully correspond to \emph{orthogonal} projective measurements, represented by 3 orthogonal Bloch vectors $\vec{a}_x$ for Alice, and by 3 orthogonal Bloch vectors $\vec{b}_y$ for Bob. 

It is convenient here to think of Alice's and Bob's measurements along the orthogonal directions $\vec{a}_x$ and $\vec{b}_y$ as the application of an adequate  local unitary operation on their respective qubit, followed by a measurement along the axes $\textsc{x}, \textsc{y}, \textsc{z}$ of their Bloch spheres\footnote{Alternatively, one can use the orthogonal directions $\vec{a}_x$ and $\vec{b}_y$ to redefine the $\textsc{x}, \textsc{y}, \textsc{z}$ axes of Alice and Bob's Bloch spheres, and hence their computational bases, and rewrite $\rho_{AB}$ in these new bases (which indeed amounts to applying local unitaries to $\rho_{AB}$).}.
Choosing random orientations for the orthogonal measurement triads $\vec{a}_x$ and $\vec{b}_y$ is equivalent to choosing random local unitary transformations to apply to the 2-qubit state $\rho_{AB}$.

In this view, the QKD protocol we are considering, with a choice of measurement among three orthogonal directions, is nothing but the entanglement-based version of the well-known 6-state protocol~\cite{bruss_6state,bechmann_gisin_6state}. Its standard security analysis can thus directly be applied. The only difference in our case here will concern its typical implementation: we shall not assume that Alice and Bob can (in the ideal case) share an entangled state with any particular symmetries adapted to their measurement bases, but instead, that their qubits undergo some uncontrolled rotations before being measured.

Note already that in the device-dependent scenario, entanglement-based protocols can readily be translated into prepare-and-measure ones~\cite{QKDreview} (whose practical implementations are typically simpler), and the following analysis would still apply.

\subsection{Device-dependent security analysis \emph{\`a la} 6-state protocol}\label{sec:DDsixState}

Following the analysis presented in Appendix~A of Ref.~\cite{QKDreview}, one can show that the asymptotic secret key rate one can extract in the 6-state protocol from the data measured (say) with the settings $x=y=3$ (corresponding to a $\sigma_\textsc{z}$ measurement, with the convention that $x,y=1$ and $x,y=2$ correspond to $\sigma_\textsc{x}$ and $\sigma_\textsc{y}$ measurements, resp.), under 1-way classical post-processing, and secure against the most general \emph{coherent attacks} (in the device-dependent scenario) is bounded by
\ba
R & \ \geq \ & 1 \, - \ H \Big[ \Big\{ \frac{1{+}E_{11}{+}E_{22}{-}E_{33}}{4}, \frac{1{+}E_{11}{-}E_{22}{+}E_{33}}{4}, \nonumber \\
&& \qquad \qquad \frac{1{-}E_{11}{+}E_{22}{+}E_{33}}{4}, \frac{1{-}E_{11}{-}E_{22}{-}E_{33}}{4} \Big\} \Big] , \label{eq:R_6st}
\ea
where $H[\{p_i\}] = -\sum_i p_i \kern0.05em \log_2 \kern0.05em p_i$ is the Shannon entropy (and $\{p_i\}$ is a length 4 vector of probabilities).

\begin{figure}%[h!]
\centering
    \includegraphics[width=8cm]{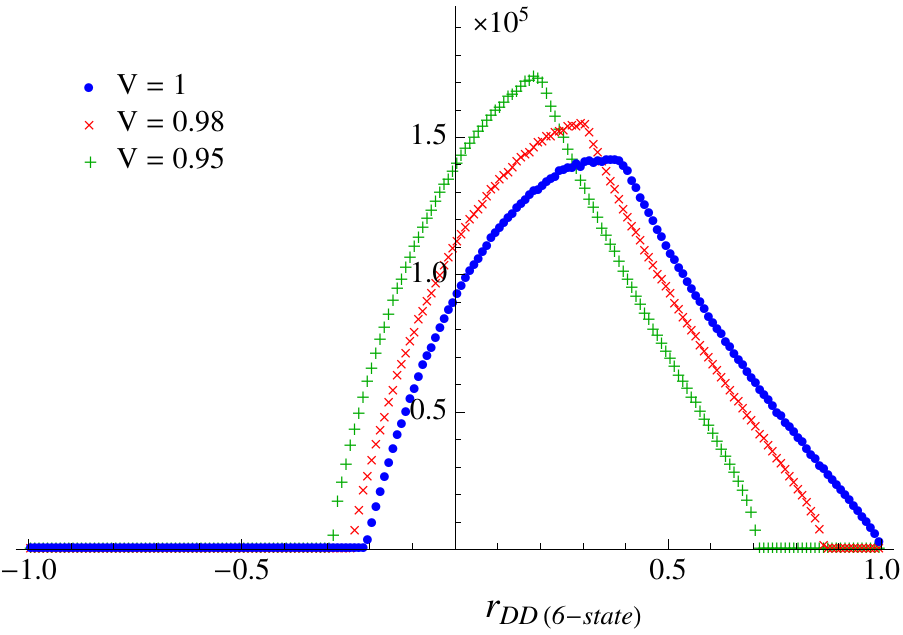}
    \caption{Estimated distribution of the bound $r_{DD (6-state)}$~(\ref{eq_rDD_6st}) on the secret key rate for Werner states of visibilities $V = 1, 0.98$ and 0.95 (in the DD scenario, following the analysis ``\emph{\`a la} 6-state protocol''), obtained as in Figures~\ref{fig_distribution_rDI_1} and~\ref{fig_distribution_rDI_2}.}
    \label{fig_distribution_rDD_6state}
\end{figure}

In our case, the association between each of Alice's 3 measurement settings and one of Bob's settings is not defined \emph{a priori}, but can be optimized so as to end up with the largest possible secret key rate---that is, we can choose the optimal permutation $\pi(\{1,2,3\}) = \{y_1, y_2, y_3\}$ of Bob's settings to be associated to Alice's settings $\{1,2,3\}$. Taking into account the fact that Eq.~\eqref{eq:R_6st} assumes a given handedness for the orientation of Alice and Bob's settings (that of $\{ \vec{\textsc{x}}, \vec{\textsc{y}}, \vec{\textsc{z}} \}$), we define
\ba
&& \hspace{-23mm} r_{DD (6-state)} \nonumber \\
&& \hspace{-22mm} = \! \max_{\stackrel{\{y_1, y_2, y_3\}}{ = \pi(\{1,2,3\})}} \! \Bigg[ 1 \, - \ H \Big[ \Big\{ \frac{1{+}\sigma_{\!\pi} (E_{1y_1}{+}E_{2y_2}{-}E_{3y_3})}{4}, \frac{1{+}\sigma_{\!\pi} (E_{1y_1}{-}E_{2y_2}{+}E_{3y_3})}{4},  \nonumber \\[-4mm]
&& \hspace{13mm} \frac{1{+}\sigma_{\!\pi} ({-}E_{1y_1}{+}E_{2y_2}{+}E_{3y_3})}{4}, \frac{1{-}\sigma_{\!\pi} (E_{1y_1}{+}E_{2y_2}{+}E_{3y_3})}{4} \Big\} \Big] \Bigg] \!, \label{eq_rDD_6st}
\ea
where $\sigma_{\!\pi} = \pm 1$ is the signature of the permutation $\pi$.
$r_{DD (6-state)}$ is thus a lower bound on the asymptotic extractable secret key rate of our device-dependent protocol, secure against coherent attacks in the limit of infinitely long keys, obtained from the standard analysis of the 6-state protocol\footnote{Note that by considering only 2 of the 3 settings of both Alice and Bob, one can follow the security analysis for the BB84 protocol~\cite{BB84} (cf e.g. the Appendix~A of~\cite{QKDreview}), and derive a simpler bound on the asymptotic secret key rate, secure against coherent attacks (and actually proven to give one-sided device independent security~\cite{TomamichelRenner,1sDI}, under the memoryless assumption), given by $r_{DD (BB84)} = \max_{x \neq x', y \neq y'} \big[ 1 - h\big(\frac{1{-}E_{xy}}{2}\big) - h\big(\frac{1{-}E_{x'y'}}{2}\big) \big]$. Numerical simulations suggest that this bound is in general only slightly lower than $r_{DD (6-state)}$~\eqref{eq_rDD_6st}, and gives comparable distributions to those of Figure~\ref{fig_distribution_rDD_6state}.}.

We again estimated the distribution of the bound $r_{DD (6-state)}$ in a similar manner as before, i.e. if Alice and Bob share (noisy) singlet states and their measurement orientations (or their local unitaries, equivalently---cf above) are chosen at random, uniformly on the Bloch sphere. The results are shown in Figure~\ref{fig_distribution_rDD_6state}. In the noiseless case ($V=1$) we found that $\sim 89.2 \%$ of our samples led to positive secret key rates $r_{DD (6-state)} > 0$. The average value of $r_{DD (6-state)}$ was found to be $\sim 0.330$; if post-selected to the cases in which $r_{DD (6-state)} > 0$, it becomes $\sim 0.379$. The maximum value of $r_{DD (6-state)}$ is 1, obtained e.g. when $\rho_{AB}$ is a pure singlet state and Alice and Bob's measurement axes are perfectly aligned (as in the standard case of the 6-state protocol).
For $V = 0.98$ and 0.95, as the key rates decrease, we still found that $\sim 84.4 \%$ and $\sim 75.6 \%$ of our samples, respectively, led to positive secret key rates $r_{DD (6-state)} > 0$.

\medskip

We note that the key rates obtained from~\eqref{eq_rDD_6st}, in the DD scenario, are typically larger than those found in the DI scenario considered in the previous section. This was expected, as the assumptions on Eve's possible attacks are more restrictive in the DD scenario. For instance, Eve cannot act on Alice and Bob's measurement apparatuses---which, in the DI scenario, indeed allows her to perform more powerful attacks~\cite{DI_NJP}. An interesting difference between the DD and DI scenarios is the optimal orientations of settings. In the DD scenario, one only aims at maximizing three of the correlators $|E_{xy}|$, and the optimal arrangement does not allow the violation of any Bell inequality; on the other hand, in the DI scenario a trade-off must be found between a large enough violation of a Bell inequality and a large enough correlator $|E_{x_{\text{raw}} y_{\text{raw}}}|$ (cf above).

\subsection{Improved device-dependent security analysis}\label{sec:DDimproved}

In the security analysis of the 6-state protocol that leads to the closed form~\eqref{eq:R_6st}, one uses the fact that an upper bound on Eve's information can be obtained, through a ``depolarization process'', by restricting oneself to Bell-diagonal states $\rho_{AB}$ (cf Appendix~A of~\cite{QKDreview}).
While this use of the symmetries of the protocol may be well adapted for standard implementations in which Alice and Bob share a common reference frame and indeed expect their state $\rho_{AB}$ to be (close to) a Bell-diagonal state, the upper bound thus obtained may in general be over-pessimistic, and it may be possible to actually derive larger bounds on the secret key rates---as we now show.

In the experimental situation we consider, Alice and Bob each repeatedly perform one out of three orthogonal qubit measurements. Their full statistics---i.e. their correlators $E_{xy}$, together with the marginal probabilities $P(a|x)$ and $P(b|y)$ (which are expected to be uniformly 1/2 for maximally entangled 2-qubit states, possibly including white noise)---then actually allow them to fully reconstruct the state $\rho_{AB}$, up to local unitary rotations, through quantum state tomography~\cite{QST}. This can be used to estimate Eve's information more tightly---e.g., in the ideal case in which the state $\rho_{AB}$ would be found to be a pure state (such as a maximally entangled state), then one can be assured that Eve is not correlated to it.

More precisely, to study the security against \emph{collective attacks}, the information potentially available to an eavesdropper can be represented by a quantum system $E$ that is correlated to Alice and Bob's system in such a way that it ``purifies'' the reconstructed state $\rho_{AB}$---i.e., one can define a purification $\ket{\psi_{ABE}}$ of $\rho_{AB}$ (a pure 3-partite state such that $\text{Tr}_E \ket{\psi_{ABE}}\!\bra{\psi_{ABE}} = \rho_{AB}$), and give the quantum system $E$ to Eve.

Let us denote by $\rho_E = \text{Tr}_{AB} \ket{\psi_{ABE}}\!\bra{\psi_{ABE}}$ Eve's partial state and by $\rho_{E|A_{x}{=}a}$ her conditional state corresponding to Alice's measurement result $A_{x}{=}a$ for the choice of setting $x$, and let us define Eve's Holevo information conditioned on Alice's outcome as
\ba
\chi(A_{x}:E) \ = \ S(\rho_E) - \sum_{a} p(A_{x}{=}a) \, S(\rho_{E|A_{x}{=}a}),
\ea
where $S$ denotes the von Neumann entropy [$S(\rho) = -\text{Tr}(\rho log_2 \rho)$]. We similarly define $\chi(B_{y}:E)$ to be Eve's Holevo information conditioned on Bob's measurement result for the choice of setting $y$.
A lower bound on the asymptotic secret key rate one can extract through 1-way post-processing from the data $A_{x_{\text{raw}}}$, $B_{y_{\text{raw}}}$ obtained from the measurement of the settings $x_{\text{raw}}$ and $y_{\text{raw}}$ is then given by the Devetak-Winter bound~\cite{DevetakWinter}
\ba
R & \ \geq \ & I(A_{x_{\text{raw}}}:B_{y_{\text{raw}}}) - \min [ \chi(A_{x_{\text{raw}}}:E), \chi(B_{y_{\text{raw}}}:E) ], \label{eq:R_DD_DevetakWinter}
\ea
where $I(A_{x_{\text{raw}}}:B_{y_{\text{raw}}})$ is the mutual information between Alice's and Bob's measurement results $A_{x_{\text{raw}}}$ and $B_{y_{\text{raw}}}$ which, after randomization of Alice and Bob's marginals (through a simultaneous random flipping of their results), is equal to $I(A_{x_{\text{raw}}}:B_{y_{\text{raw}}}) = 1 - h\big[ \frac{1{-}E_{x_{\text{raw}} y_{\text{raw}}}}{2} \big]$.
The bound~\eqref{eq:R_DD_DevetakWinter} ensures the security of the secret key against collective attacks (in the limit of infinitely long keys); using a de Finetti type of argument, one can show that the same secret key rate is also secure against \emph{coherent attacks}~\cite{Renner_deFinetti_argument}.

\begin{figure}%[h!]
\centering
    \includegraphics[width=8cm]{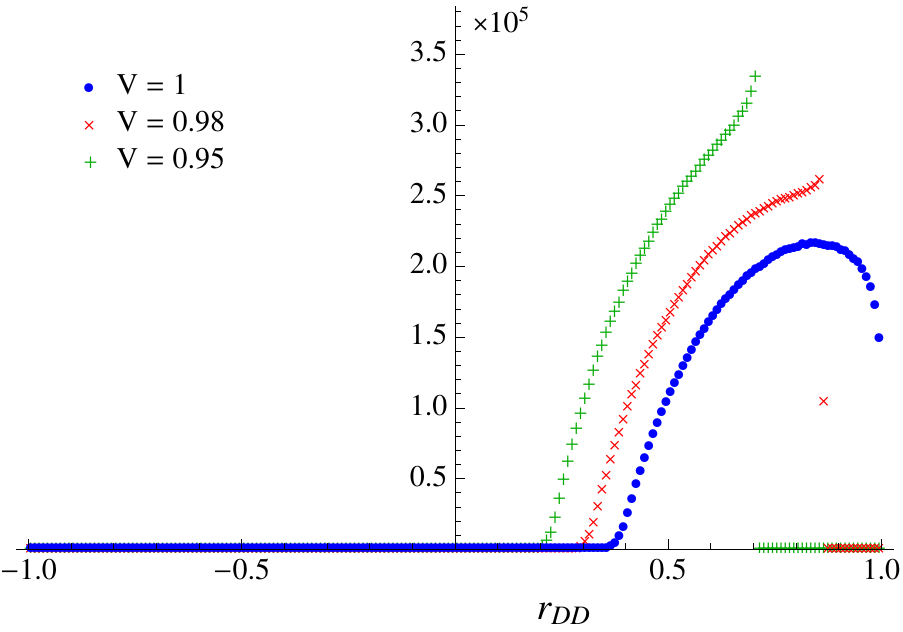}
    \caption{Estimated distribution of the bound $r_{DD}$~(\ref{eq_rDD_opt}) on the secret key rate for Werner states of visibilities $V = 1, 0.98$ and 0.95 (in the DD scenario, improving on the standard analysis for the 6-state protocol), obtained as in Figures~\ref{fig_distribution_rDI_1}--\ref{fig_distribution_rDD_6state}.}
    \label{fig_distribution_rDD_opt}
\end{figure}

In our case, Alice and Bob still have the possibility to choose the settings from which they will attempt to extract a secret key. Let us accordingly define
\ba
\hspace{-15mm}
r_{DD} \ = \ \max_{x_{\text{raw}},y_{\text{raw}}} \ \Bigg[ 1 \, - \ H \Big[ \frac{1{-}E_{x_{\text{raw}} y_{\text{raw}}}}{2} \Big] - \min \big[ \chi(A_{x_{\text{raw}}}:E), \chi(B_{y_{\text{raw}}}:E) \big] \Bigg], \label{eq_rDD_opt} 
\ea
If $r_{DD}$ is found to be non-negative, then Alice and Bob will actually be able to extract a secret key with a rate (at least) $r_{DD}$---which is larger than the previous bound $r_{DD (6-state)}$~\eqref{eq_rDD_6st}.

\medskip

In the ideal case in which Alice and Bob find that they share noiseless singlet states, the state $\rho_{AB}$ is pure. This implies in particular that Eve's Holevo information is null: $\chi(A_{x_{\text{raw}}}:E) =  \chi(B_{y_{\text{raw}}}:E) = 0$. The bound $r_{DD}$~\eqref{eq_rDD_opt} on the secret key rate then just depends on the largest correlator (in absolute value) $E_{xy} = - \vec a_x \cdot \vec b_y$ observed by Alice and Bob. One can show in that case that if Alice's and Bob's 3 measurements settings are orthogonal, this largest correlator is necessarily greater than $\frac{2}{3}$ (obtained if all scalar products $\vec a_x \cdot \vec b_y$ are either $\pm \frac{2}{3}$ or $\pm \frac{1}{3}$), and hence $r_{DD} \geq 1 - h\big( \frac{1}{6} \big) \simeq 0.350 > 0$; on the other hand, the maximum value 1 of $r_{DD}$ is attained if any two of Alice and Bob's settings are aligned.
As in the previous cases, we estimated the distribution of $r_{DD}$ for randomly chosen orientations for Alice and Bob's measurement triads; see Figure~\ref{fig_distribution_rDD_opt}. Its average value was found to be $\sim 0.745$.

If Alice and Bob determine that they find a noisy Werner state with $V < 1$, they calculate Eve's Holevo information to be (for all $x,y$) $\chi(A_x:E) = \chi(B_y:E) = H \big[ \{ \frac{1+3V}{4}, \frac{1-V}{4}, \frac{1-V}{4}, \frac{1-V}{4} \} \big] - h \big[ \frac{1-V}{2} \big]$. The distribution of $r_{DD}$, estimated as before, is also shown on Figure~\ref{fig_distribution_rDD_opt} for $V = 0.98$ and $V = 0.95$. In both cases we always find positive bounds $r_{DD}$ on the secret key rates (in fact, $r_{DD}$ is always positive for $V \gtrsim 0.875$).

\section{Proof-of-principle experiments}

To demonstrate the practical relevance of our theoretical discussions, we performed a proof-of-principle demonstration of QKD using the four security analyses discussed above. In our experiment Alice generated a sequence of pairs of polarization entangled photons and sent one photon of each pair to Bob via a channel with an unknown polarization transformation. Both parties projectively measured the polarization state of their photon in one of three mutually unbiased bases. Neither Alice nor Bob attempted to align their measurement devices, as we do not want to assume that they share a common reference frame. After collecting sufficient data on each pair of projectors---giving the 9 correlators $E_{ij}$ as described above and allowing for the tomography of the quantum state shared by Alice and Bob (for the calculation of $r_{DD}$)---asymptotic secret key rates from each of the above analyses were calculated. Note that our demonstration is proof-of-principle only as we do not randomly select bases nor perform the required error correction or privacy amplification, which is required to generate actual secret keys (and which would require a rigorous finite-key analysis, which would go beyond the scope of this paper). Nor do we close the detection loophole as necessary to generate key for device-independent QKD.

\subsection{Experimental setup}

\begin{figure}%[h!]
\centering
    \includegraphics[width=11cm]{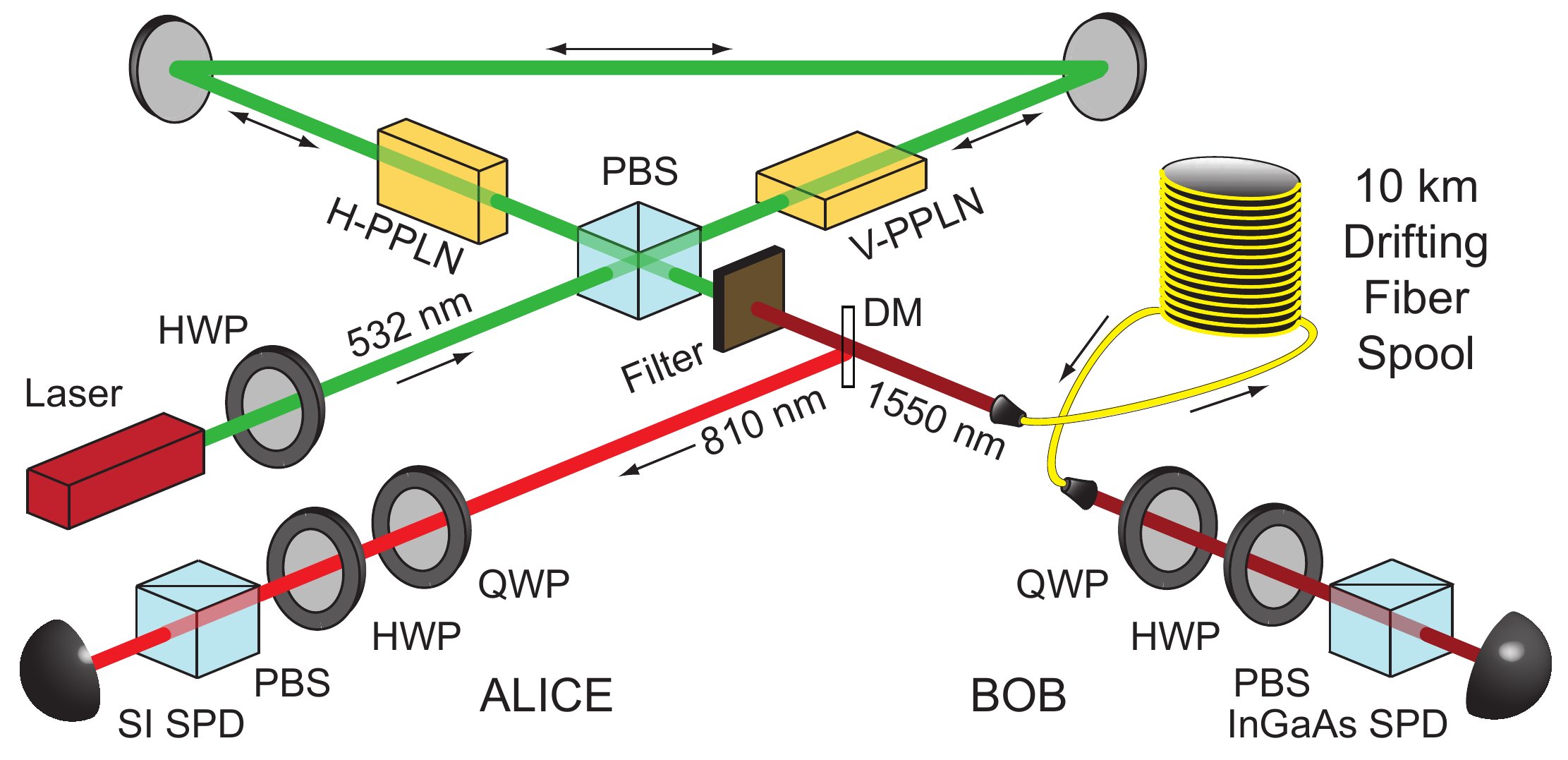}
    \caption{Experimental setup. A horizontally polarized, pulsed ($50$~ps), $532$~nm wavelength laser beam is rotated to diagonal polarization via a half wave-plate (HWP), is then split by a polarizing beam splitter (PBS) and travels both clockwise and counter-clockwise through a polarization Sagnac interferometer. The interferometer contains two type-I, spontaneous parametric down-conversion (SPDC), periodically-poled lithium niobate (PPLN) crystals configured to produce collinear, non-degenerate, 810/1550 nm wavelength photon pairs. The clockwise-travelling, vertically polarized (counter-clockwise travelling, horizontally polarized) pump light passes through the first crystal without interaction (as SPDC is polarization dependent) and may down-convert in the V-PPLN crystal (H-PPLN) to produce two horizontally (vertically) polarized photons. After exiting the interferometer, the remaining pump light is filtered out using a high-pass filter and the entangled photons are separated on a dichroic mirror (DM) and sent to qubit analyzers consisting of waveplates, a PBS and single-photon detectors.}
    \label{fig_ExpSetup}
\end{figure}

Fig.~\ref{fig_ExpSetup} shows a schematic of our experiments. Alice holds a source of polarization-entangled qubits and a qubit analyzer (details below). Her entanglement source is based on two spontaneous parametric downconversion (SPDC) crystals in a polarizing Sagnac interferometer, described and characterized previously in~\cite{Stuart12,Stuart13}. Diagonally polarized pump light from a pulsed $532$~nm wavelength laser is placed in a superposition of traveling clockwise (CW) and counter-clockwise (CCW) around the Sagnac interferometer. In the CCW path, vertically polarized pump light passes unaffected through the first SPDC crystal, which is oriented to down-convert horizontally polarized pump light, and then produces pairs of vertically polarized photons in the second SPDC crystal. Similarly, the CW path produces pairs of horizontally polarized photons. Each pair consists of one photon at around $810$~nm wavelength and one photon at around $1550$~nm. By recombining the two paths at the output of the interferometer, and keeping pump powers sufficiently low, Alice generates a two-qubit state close to the $\ket{\Phi^+}$ Bell state\footnote{Note that all maximally entangled 2-qubit states (such as $\ket{\Phi^+}$) are equivalent, up to a local unitary transformation, to the singlet state $\ket{\Psi^-}$ considered in the previous sections.}. Performing quantum state tomography based on a maximum likelihood optimization~\cite{QST} with the source revealed an average tangle of ${\cal T} = 0.85 \pm 0.02$ (note that ${\cal T} = 1$ implies a maximally entangled state and that we observed the tangle to oscillate between $0.82$ and $0.88$ over the course of the experiment); this value of the tangle corresponds, for an ideal Werner state, to an average visibility of about $V = 0.95 \pm 0.01$.

During experiments, Alice separates the two entangled photons with a dichroic mirror and measures the $810$~nm photon directly with her qubit analyzer, which consists of waveplates, a polarizing beam splitter (PBS) and a free-running silicon avalanche photo-diode (APD). Alice also sends the $1550$~nm photon to Bob via a 10~km fiber spool with approx.~$6$~dB loss, which serves as the quantum channel with unknown and varying polarization transformation, and, in parallel, generates an electronic signal to inform Bob of the incoming photon. Bob then projectively measures the photon with his own qubit analyzer, also consisting of waveplates, a PBS and a gated InGaAs APD. Measurement results from both Alice and Bob are recorded on the same PC for analysis.

\subsection{Experimental results}

To demonstrate the feasibility of QKD without a shared reference frame with our setup, we performed two experiments. In both experiments Alice and Bob collected statistics on one of the nine correlators for two minutes and then either Alice or Bob would change measurement settings. Hence, $18$~minutes were required to collect statistics on all nine correlators. In the first experiment, Alice and Bob cycled through the measurement settings for nearly three hours while the polarization transformation of the fiber spool was allowed to freely drift, which generated nine complete iterations through the measurements of the correlators. For our second experiment, we inserted three waveplates into the channel connecting Alice and Bob to randomly vary the polarization transformation and measured the nine correlators for each transformation.

In the first (free-drifting) experiment we analyzed the nearly three hours of data with a sliding $18$~minute window: We first analyzed the nine correlators and performed state tomography with the data within a window beginning at time $t = 0$ and going to $t = 18$~min. We then repeatedly stepped the window forward by $2$~minutes, yielding a total of $73$ sets of data (e.g. the second data set is between $t = 2$~min and $t = 20$~min, etc.), and analyzed each set independently. For each data set we calculated the asymptotic secret key rates $r_{DI_1}$, $r_{DI_2}$, $r_{DD (6-state)}$ and $r_{DD}$ one could extract in each of the 4 scenarios, according to Eqs.~\eqref{eq_rDI_1}, \eqref{eq_rDI_2}, \eqref{eq_rDD_6st} and~\eqref{eq_rDD_opt}, respectively.
In order to illustrate the role of the CHSH violation in the DI scenario and the importance of having large correlators, we also calculated the maximal CHSH value\footnote{Note that we do not claim that this maximal $S_{max}$ value is necessarily the one that grants the maximum secret key rate (as the latter also depends on the correlator $E_{x_{\text{raw}} y_{\text{raw}}}$; cf Eqs~\eqref{eq_rDI_1} and \eqref{eq_rDI_2}). However, as having a large value of $S_{max} (> 2)$ is a necessary condition for positive DI secret key rates, we will use the $S_{max}$ value as an indicator of the ability to generate key in the DI scenario.} $S_{max} = \max_{x,x',y,y'} S_{xx'yy'}$, and the largest sum of 3 correlators defined as $C_{max} = \max_{x \neq x' \neq x'', y \neq y' \neq y''} |E_{xy}| + |E_{x'y'}| + |E_{x''y''}|$.

\begin{figure}%[h!]
\centering
    \includegraphics[width=6.5cm]{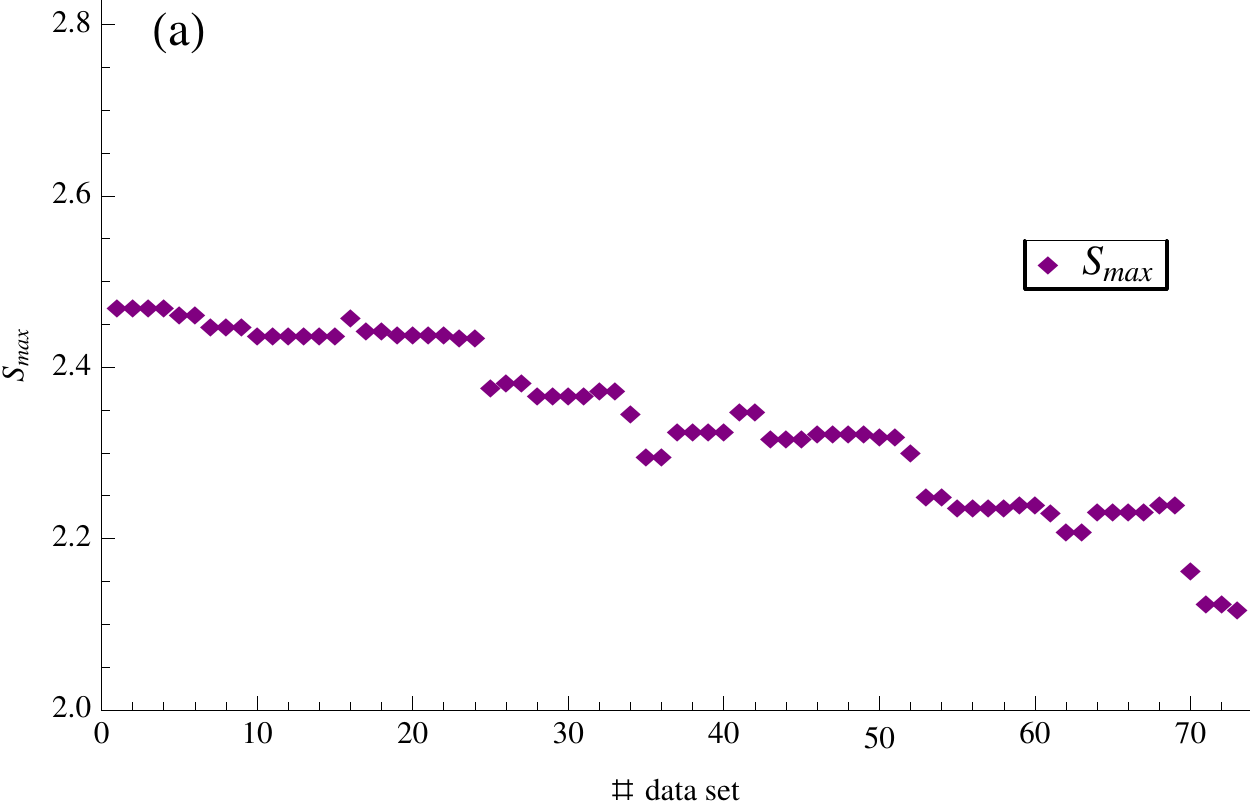} $\qquad$
    \includegraphics[width=6.5cm]{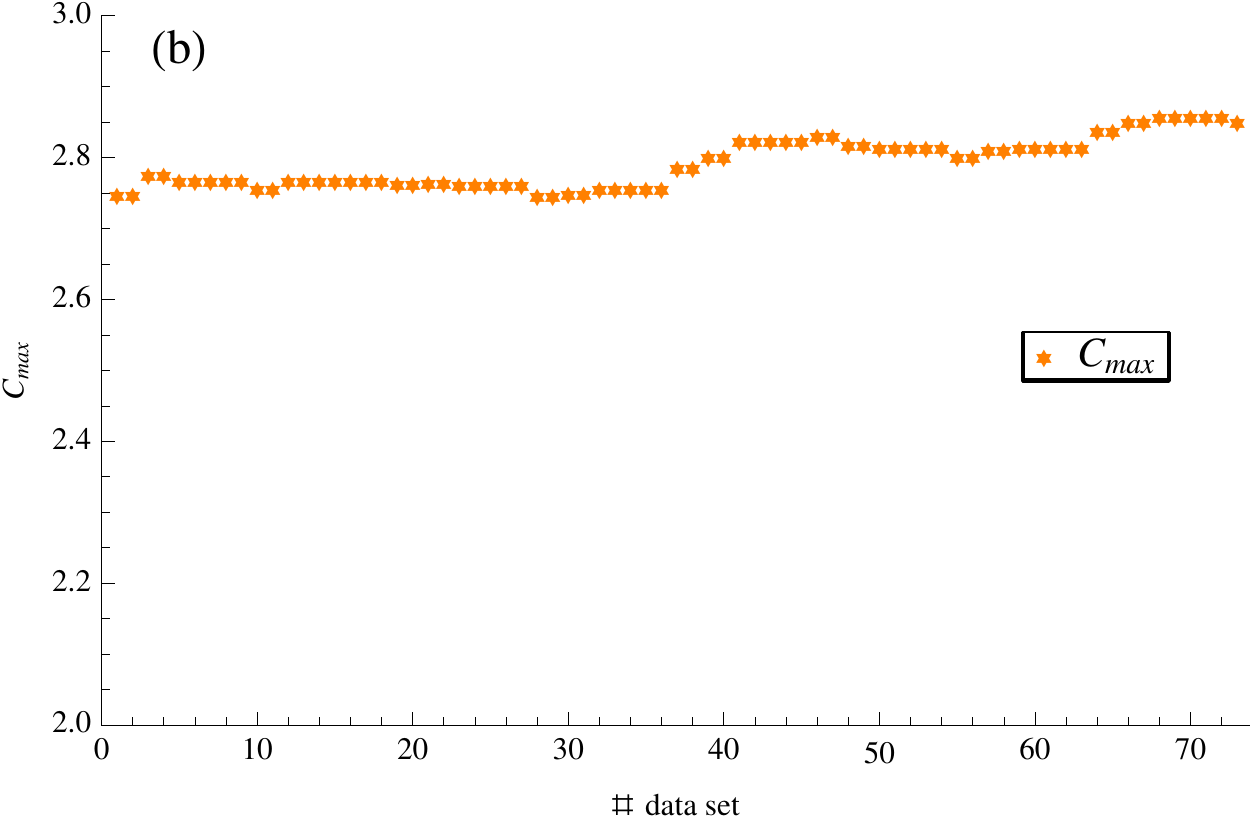} \\[5mm]
    \includegraphics[width=8cm]{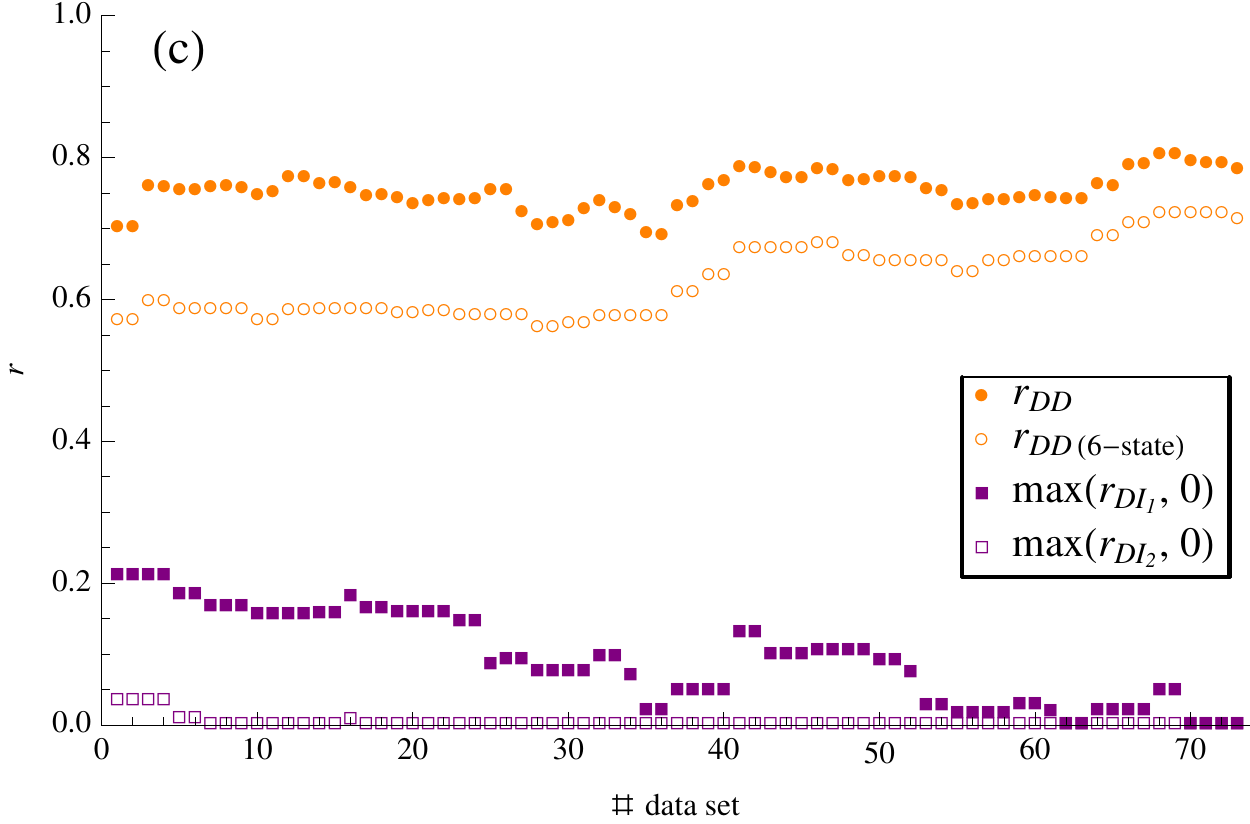}
    \caption{Results from our free-drifting experiment, as a function of time (for each of our 72 data sets; see main text). (a--b)~Our figures of merit: the maximal CHSH value $S_{max}$ and the maximal sum of correlators $C_{max}$, as defined in the main text. (c)~Asymptotic secret key rates $r_{DI_1}$, $r_{DI_2}$, $r_{DD(6-state)}$ and $r_{DD}$ corresponding to each of the 4 scenarios studied in Sections~\ref{sec:DI} and~\ref{sec:DD}.}
    \label{fig:FreeDrift}
\end{figure}

These results are presented in Fig.~\ref{fig:FreeDrift}. Initially, and by chance, the channel transformation turned out to be such that a reasonably high parameter $S_{max}$ was found, favoring device-independent QKD, but over the course of the experiment, the channel transformation slowly drifted close to a point where Alice's and Bob's measurement bases were aligned. At this point we observed 3 high correlators (i.e. a large value for $C_{max}$) and a low parameter $S_{max}$, which favours device-dependent QKD. Indeed, when one examines the key rates as a function of time (i.e. window position) one observes steadily decreasing device-independent key rates (in fact, $r_{DI_2}$ quickly falls to zero, whereas $r_{DI_1}$ remains positive for longer) and steadily increasing device-dependent key rates.

These observations align with our discussion at the end of Sec.~\ref{sec:DDsixState} in that the alignment of bases optimal for device-dependent and device-independent QKD are different. All our protocols require a source with a high degree of entanglement (characterized, for instance, by a high visibility or high tangle). However, device-dependent QKD is optimal when Alice's and Bob's measurement bases are well aligned such that one finds three large-valued correlators with which to generate key, which minimizes key reduction due to error correction. On the other hand, in device-independent QKD one requires a set of four correlators that generate a high $S$-parameter (to minimize the bound on Eve's information and thus key reduction during privacy amplification) with one correlator $E_{x_{\text{raw}} y_{\text{raw}}}$ being large so that error correction is minimal, which are conflicting requirements. This conflicting nature is indeed illustrated in Fig.~\ref{fig:FreeDrift}, where one observes that $S_{max}$ decreases in time while $C$ increases---just as $r_{DI_1}$ and $r_{DI_2}$ decrease as $r_{DD(6-state)}$ and $r_{DD}$ increase.

Furthermore, the difference between secret key rates granted by the two device-dependent analyses, $r_{DD(6-state)}$ and $r_{DD}$, are apparent in Fig.~\ref{fig:FreeDrift}(c). The difference between these techniques, as discussed in Sec.~\ref{sec:DDimproved}, is how one bounds an eavesdropper's information: $r_{DD}$ uses the Holevo information based on a reconstruction of the density matrix $\rho_{AB}$ while $r_{DD(6-state)}$ uses three correlators. If the quantum state that Alice and Bob share contains a high tangle, then Eve's Holevo information will be low and thus $r_{DD}$ mainly depends on the strength of the correlator used to generate key.  On the other hand, the privacy bound for $r_{DD(6-state)}$ does depend on alignment as three high correlators are needed for key generation. We see that in our results: first, $r_{DD}$ always outperforms~$r_{DD(6-state)}$ (as mentioned in Sec.~\ref{sec:DDimproved}) and, second, the difference between the two key rates is largest initially (when bases are the most misaligned) and slowly decreases as Alice's and Bob's bases begin to align.

\medskip

\begin{figure}%[h!]
\centering
    \includegraphics[width=6.5cm]{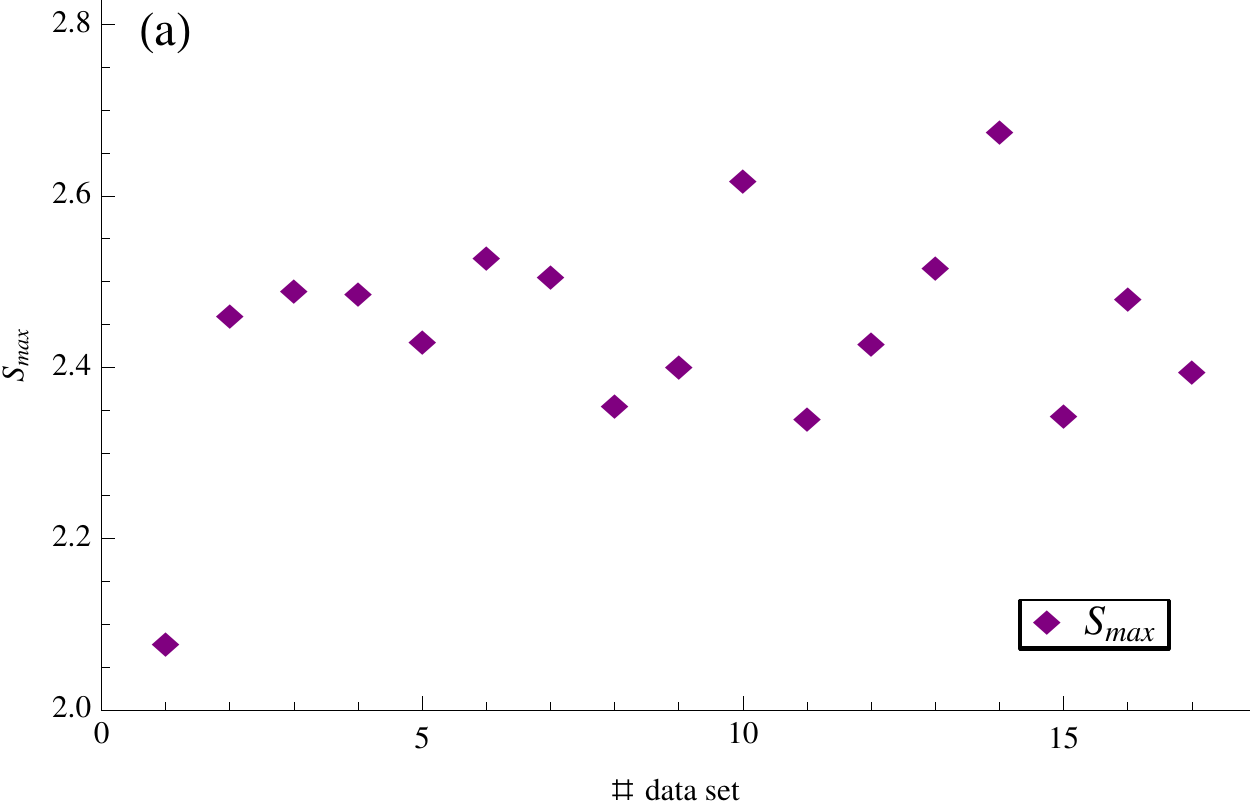} $\qquad$
    \includegraphics[width=6.5cm]{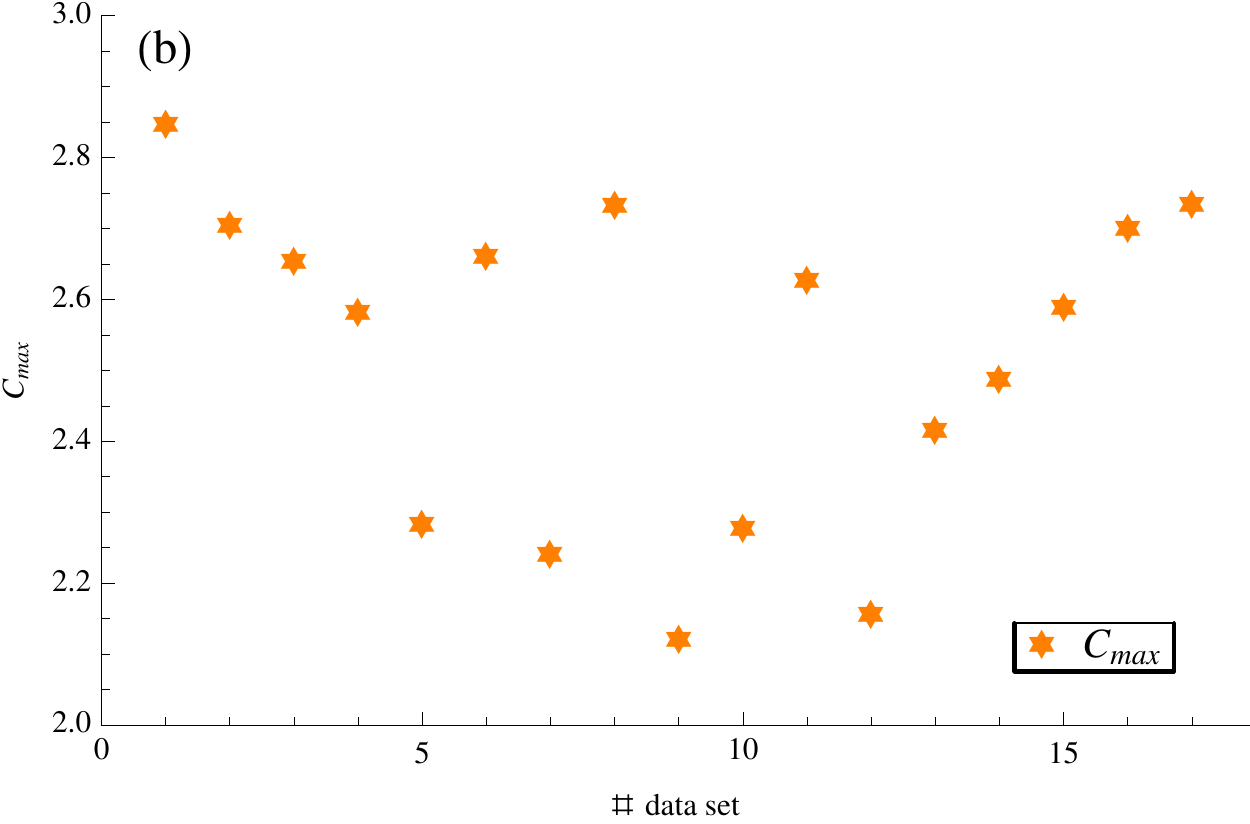} \\[5mm]
    \includegraphics[width=8cm]{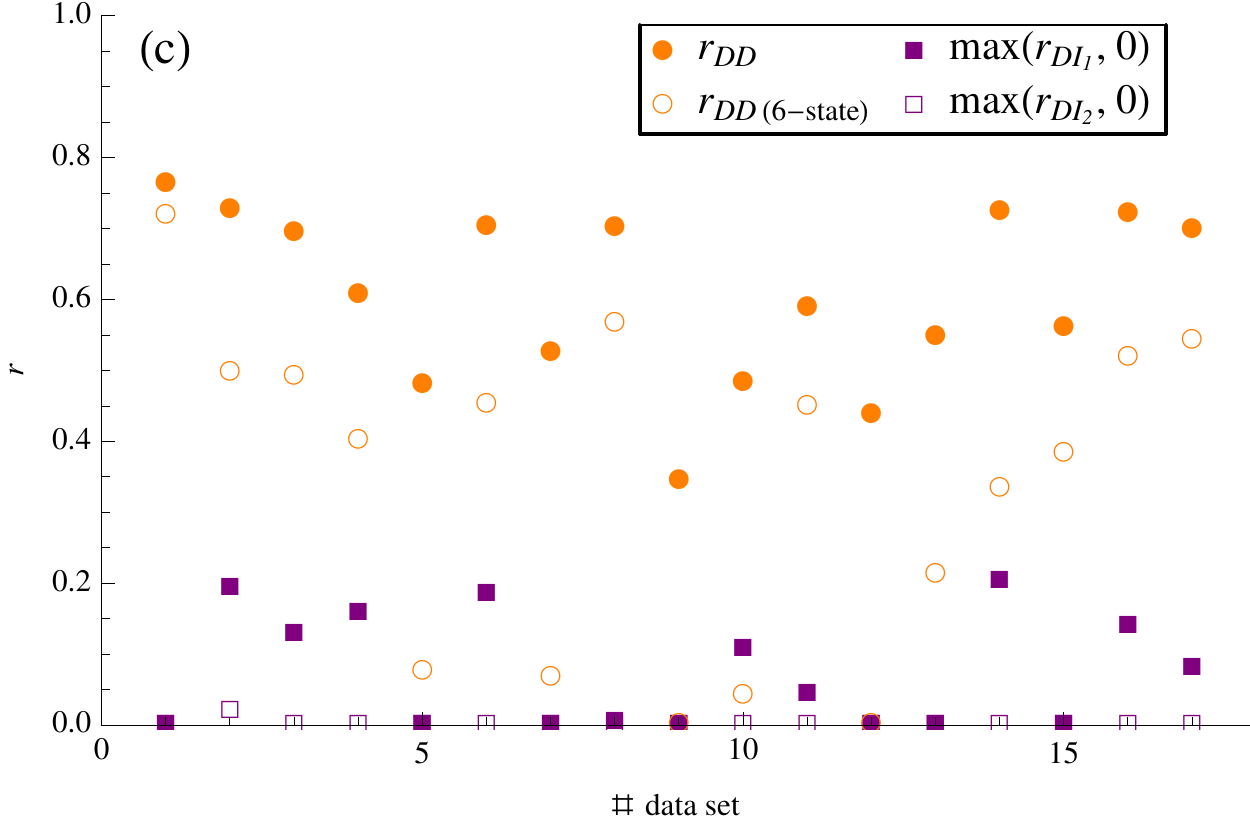}
    \caption{Results from our randomized polarization transformation experiment, for each of our 17 experimental runs (see main text). (a--c) as in Figure~\ref{fig:FreeDrift}.}
    \label{fig:RandDrift}
\end{figure}

As mentioned above, for our second experiment, we inserted three waveplates into the channel connecting Alice and Bob to randomly vary the polarization transformation.  All nine correlators were measured $17$ times, and state tomography was performed independently each time. Before each iteration the waveplates were re-positioned based on randomly generated numbers, thus generating a random channel transformation (note that the fiber's own transformation continued to drift as above). In Fig.~\ref{fig:RandDrift} we present the results. We again plot the figures of merit $S_{max}$ and $C_{max}$ for each of the $17$ measurements, as well as the derived asymptotic secret key rates $r_{DI_1}$, $r_{DI_2}$, $r_{DD (6-state)}$ and $r_{DD}$. For device-independent QKD, we found positive secret key rates for $r_{DI_1}$ in $10$ out of $17$ measurements (i.e. with probability of $59$\%) and $1$ out of $17$ measurements (i.e. $6$\%) for $r_{DI_2}$. For device-dependent QKD, we found positive secret key rates for $r_{DD}$ in $17$ of $17$ measurements (i.e. $100$\%) and in $15$ of $17$ measurements (i.e. $88$\%) for $r_{DD(6-state)}$.
Although the size of our experimental sample is too small to really be statistically significant, our observations appear to agree reasonably well with the predictions from the numerical simulations above, assuming a source of entangled Werner states of visibility $V$ slightly larger than 0.95, i.e. a tangle of $\simeq 0.856$---in agreement with the measured tangle of the state, which was found to oscillate from $0.82$ to $0.88$.

Lastly, we again point out that a consistently high tangle, which our experiment maintained (up to the oscillations), is required but not sufficient to generate positive secret key rates. A high-quality source does not guarantee the high $S$-parameter needed for device-independent QKD, nor the high correlators needed for device-dependent QKD. An appropriate channel transformation is also required.

\section{Discussion}

We have presented a practical QKD setup in which the requirement of a common reference frame can be completely dispensed with. A proof-of-principle demonstration of our protocols, which covers both the usual device-dependent case and the device-independent case, was performed over $10$~km of spooled fiber. Specifically, we have shown that a secret key can in principle be established, considering both a freely drifting spool and randomly chosen transformations, even in the device-independent case (assuming fair sampling, and infinitely long keys).

We believe that the present ideas have potential to find applications in future long-distance quantum communication protocols, in particular in situations where the amount time available to perform the protocol is severely constrained, e.g. in satelite based quantum communications.  The present results should be considered as a proof-of-principle experiment, and several technical improvements are required, such as implementing random choices of measurement settings, and a finite-key security analysis~\cite{Scarani08}. For the device-independent approach, an essential step is to close the detection loophole, which has recently been achieved in fully optical systems~\cite{Giustina13,Christensen13}. Finally, another challenge consists in devising efficient error-correction protocols for high error rates, as our protocols typically lead to higher error rates compared to the standard approach in which the parties share a common reference frame.

\bigskip

{\bf \emph{Acknowledgements.}} This work was supported by a UQ Postdoctoral Research Fellowship, the Swiss National Science Foundation (grant PP00P2\_138917), the EU project DIQIP, the National Sciences and Engineering Research Council of Canada (NSERC), Alberta Innovates Technology Futures (AITF), the Canadian Foundation for Innovation (CFI), Alberta Advanced Education and Technology (AAET) and by the Killam Trusts.

\end{document}